\documentclass[fleqn,12pt]{wlscirep}

\usepackage{color}
\usepackage[normalem]{ulem}
\usepackage{amsmath,amssymb,amsfonts}%

\newtheorem{definition}{Definition}

\newtheorem{example}{Example}
\newcommand{\EE}{\mathrm{E}}

\newcommand{\VV}{\mathrm{Var}}

\title{Measuring multisensory integration in reaction time: the relative entropy approach}

\author[1,*]{Hans Colonius}
\author[1]{Adele Diederich}

\affil[1]{Carl von Ossietzky Universit\"{a}t Oldenburg, Department of Psychology, Oldenburg, 26129, Germany}

\affil[*]{hans.colonius@uni-oldenburg.de}


\begin{abstract} 
A classic definition of \textit{multisensory integration} (MI) has been proposed as ``the presence of a (statistically) significant change in the response to a cross-modal stimulus complex compared to unimodal stimuli''. However, this general definition did not result in a broad consensus on how to quantify the amount of MI in the context of reaction time (RT). In this brief note, we argue that numeric measures of reaction times that only involve mean or median RTs do not uncover the information required to fully  assess the effect of multisensory integration. We suggest instead novel measures that include the entire RT distributions functions. The central role is played by \textit{relative entropy} (aka \textit{Kullback-Leibler divergence}), a statistical  concept in information theory, statistics, and machine learning to measure the (non-symmetric) distance between probability distributions.  We provide a number of theoretical examples, but empirical applications and statistical testing are postponed to later study. 
\end{abstract}

\begin{document}
\flushbottom
\maketitle
\thispagestyle{empty}

The study of how information from different sensory modalities is merged to produce a unified percept is an important topic in many research fields including the behavioral sciences. A pragmatic definition of \textit{multisensory integration} (MI) as ``\textit{the presence of a (statistically) significant change in the response to a cross-modal stimulus complex compared to unimodal stimuli}'' has been proposed in \cite{stein2009}. In the realm of reaction time (RT) measures for MI, this amounts to comparing the time e.g., to detect a visual-auditory stimulus, to the time to detect to a unisensory, visual or auditory, stimulus. The study of cross-modal interaction effects in RTs  goes back more than 100 years\cite{Todd1912}, and has generated a huge number of studies (see reviews \cite{hershenson1962,welch1986,rach2011}).  

Here, we argue that current MI measures for reaction times based solely on parameters of central tendency, like means or medians,  do not fully reveal the information available to assess effects of multisensory integration. 
We suggest a novel approach to quantifying MI that involves  the entire RT distributions,  with no underlying parametric modeling assumptions. 
The basic idea is to measure, in a sense to be specified below, how ``far away'' the cross-modal RT distribution is from the unimodal RT distributions. A central role is played by the concept of \textit{relative entropy}  (aka \textit{Kullback-Leibler divergence}), a statistical  concept in information theory, statistics, and machine learning to measure the (oriented, i.e. non-symmetric) distance between probability distributions\cite{CoverThomas1991}. The new measures is illustrated by some theoretical examples. Empirical applications including simulation and testing have to be postponed to future study.

First, we introduce the traditional measure of multisensory integration for RTs and point out its shortcomings due to being based on means or medians only. Then, the notion of relative entropy is defined and some of its properties are presented. Two sets of examples follow: (i)~defining crossmodal response enhancement (CRE) for different statistical distributions (exponential, normal, and lognormal) as an alternative to the classical one based on means (expected values) only; (ii)~using relative entropy to quantify by how much a model's prediction deviates from observed data, thereby replacing a well-known measure relating to the race model for bisensory RTs.

\subsection*{Response enhancement in redundant signals paradigm: the traditional measure}
In the \emph{redundant signals paradigm}, also known as \emph{divided attention paradigm}, a participant is instructed to respond as soon as a uni- or cross-modal signal occurs.  A traditional measure of cross-modal response enhancement (CRE) in RTs is defined as\cite{rach2011}
\begin{equation}\label{crert+}
	\mathrm{CRE}_{RT}=  \frac{\min\{\EE RT_V,\EE RT_A\}-\EE RT_{V A}}{\min\{\EE RT_V,\EE RT_A\}}\times 100.
\end{equation}
Here, $\EE$ stands for expected (mean) value, but the median is often used instead as well. The numerator compares the faster of the unisensory RTs (here, visual or auditory) to the cross-modal (visual-auditory) RT, and the denominator and multiplication factor simply serve to standardize the measure. Thus, $\mathrm{CRE}_{RT}$ expresses multisensory enhancement or inhibition as a proportion of the faster unisensory response. For example, $\mathrm{CRE}_{RT}=10$  means that mean RT to the visual-auditory stimulus is 10\% faster than the faster of the expected RTs to the unimodal visual or auditory stimuli. For  simplicity, we neglect occurrence of erroneous responses, like failure to detect a stimulus. 

Measure $\mathrm{CRE}_{RT}$ is a simple way of quantifying MI  that is amenable to standard statistical testing. However,  it does not take into account that integrating information from different modalities may also affect other, more fine-grained aspects of the associated RT distributions. For example, one possible result of integrating information might be that short RTs become more frequent while long RTs tend to be even longer, leaving the difference between uni- and cross-modal mean RTs more or less invariant. Because stimulus detection is conceived of as a stochastic event generating some random variability in information accumulation, the way this variability is modified under cross-modal stimulation may yield important insights into the integration process itself\cite{otto2013}. 
\section*{Multisensory integration measures based on relative entropy}\label{KLD}
All available  information about the MI process is contained in how the multisensory RT distribution differs from the unimodal RT distributions. Thus, an MI measure should be some function of this difference. We identify two issues: first, how should this difference be formally defined? Second, how should the two unisensory RT distributions be combined to enter into that expression?

Recall that a metric $d$ on a set $S$ is defined as a function $d\colon S \times S \to\mathbb{R}$ ($\mathbb{R}$ the set of real numbers) such that, for all $x,y,z \in S$, (i)~$d(x,y)\ge 0$ (\textit{non-negativity}); (ii)~$d(x,y)=0$ if and only if $x=y$; (iii)~$d(x,y)=d(y,x)$ (\textit{symmetry}); and (iv)~$d(x,y) \le d(x,z) + d(z,y)$ (\textit{triangle inequality}).

There are many ways to define  a metric  on a set of probability\cite{deza2009}. However, it turns out that not all properties of a metric  are actually needed for our approach. We want a measure that quantifies how much the unimodal distributions have to be ``modified'' in order to attain the cross-modal distribution; thus, neither symmetry nor the triangle inequality are required. This suggests using the following concept: 
\begin{definition}
	The  \emph{relative entropy (RE) } between two probability mass functions $p(x)$ and $q(x)$  is defined as
	\begin{align}
		D(p||q)&= \sum_{x \in X} p(x) \log \frac{p(x)}{q(x)}\\
		&=\EE_p  \log \frac{p(X)}{q(X)}.\label{exp}
	\end{align}
\end{definition}
Here,  $X$ is a discrete real-valued random variable and Equation~(\ref{exp}) means that $D(p||q)$ equals the expected value of random variable $\log \frac{p(X)}{q(X)}$ with respect to probability mass function $p(x)$. We use the convention that $ 0 \log \frac{0}{q} = 0$ and $p \log \frac{p}{0}=0$. Relative entropy is also known as \emph{Kullback-Leibler Divergence} (KLD). Relative entropy for continuous random variables with probability density functions (pdf) $f$ and $g$ is defined as
\[ D(f||g) = \int f \log \frac{f}{g}. \]
$D(p||q)$ can be interpreted as measuring how well the ``target'' distribution $p(x)$ is approximated by $q(x)$; it plays an important role in several fields including information theory, statistical physics, neural networks, and Bayesian statistics\cite{KL1951,CoverThomas1991,MacKay2003}. 
Relevant properties  for our purposes are:
\begin{enumerate}
	\item $D(p||q)=0$ if and only if $p=q$   (\emph{self-identification})
	\item $D(p||q)\ge 0$ for all $p,q$  (\emph{non-negativity})
\end{enumerate}
Non-negativity, also known as \textit{Gibb's inequality} or \textit{information inequality}, follows from Jensen's inequality (for proofs, see the above references). 

\subsection*{Defining a measure of MI based on relative entropy}
The measure of MI based takes the cross-modal pdf, $f_{VA}(t)$,  as  the ``target'' function $p(x)$ (or $f$) and unimodal pdfs $f_V, f_A$   (indexes $VA,V$, and $A$ here stand again for visual-auditory crossmodal and unimodal conditions). Without adding any modeling assumption, we take the smaller of the Kullback-Leibler divergences (KLD) with respect to the unisensory distributions  to define  (note that we are using the shorthand KLD from now on):
\begin{definition}
	\begin{equation}\label{creKLD}
	\mathrm{CRE}_{KLD} = \min\{D(f_{VA} || f_V), D(f_{VA} || f_A)\}.
	\end{equation}
\end{definition}
This is somewhat analogous to the traditional measure $\mathrm{CRE}_{RT}$ of Equation~(\ref{crert+}). It equals zero if $f_{VA}= f_V$ or $f_{VA}=f_A$. Note that KLD values can take large values going towards infinity. In order to make $\mathrm{CRE}_{KLD}$ values from different data sets comparable, a standardization like in Equation~(\ref{crert+}) would be desirable, but it seems not obvious how to do this.

%
%
\begin{example}[Exponential]
	 We assume that $f_{VA}	, f_V, f_A$ are exponential distributions with parameters $\lambda_{VA}, \lambda_V, \lambda_A$, respectively, and let $\lambda_{VA}> \lambda_A>\lambda_V>0$. Then
	\begin{align*}
		D(f_{VA}|| f_V) &= \int_{0}^{\infty} \lambda_{VA} \exp(-\lambda_{VA}\, t) \log\left[\frac{\lambda_{VA} \exp(-\lambda_{VA} \,t)}{\lambda_V \exp(-\lambda_V\, t)}\right] dt\\
		&= \log \frac{\lambda_{VA}}{\lambda_V} + \frac{\lambda_V}{\lambda_{VA}} -1,
	\end{align*}
	so that 
	\begin{align*}
	\mathrm{CRE}_{KLD} &=  \min\{D(f_{VA} || f_V), D(f_{VA} || f_A)\}\\
		&= \min\{ \log \frac{\lambda_{VA}}{\lambda_V} + \frac{\lambda_V}{\lambda_{VA}} -1,\log \frac{\lambda_{VA}}{\lambda_A} + \frac{\lambda_A}{\lambda_{VA}} -1\}\\
		&=  \log \frac{\lambda_{VA}}{\lambda_V} + \frac{\lambda_V}{\lambda_{VA}} -1.
	\end{align*}
	The last equality follows from assuming $ \lambda_A>\lambda_V$. With $(\lambda_{VA}/\lambda_V) \to \infty$, that is, if the effect of MI increases without bound, then $\mathrm{CRE}_{KLD}\to\infty$ as well. Note that the  exponential is not a plausible RT distribution and is presented here for illustration only.
\end{example}
\begin{example}[Normal]
	The density of the normal distribution of a real-valued random variable $X$ is 
	\[f(x;\mu_x, \sigma_x) = \frac{1}{\sigma \sqrt{2\pi}}\exp\left(- \frac{(x-\mu)^2}{2 \sigma^2}\right)\]
	with $\mu_x \in (-\infty, +\infty)$ and $\sigma_x >0$, abbreviated as $X\sim \mathcal{N} (\mu_x,\sigma^2_x)$. The KLD for two random variables $X$ and $Y$ with $Y\sim \mathcal{N} (\mu_y,\sigma^2_y)$ with densities $f_x, f_y$, respectively, is known to equal
	\begin{align}
		D(f_x||f_y)= \frac{1}{2} \left[             \frac{(\mu_x-\mu_y)^2}{\sigma^2_y} + \frac{\sigma^2_x}{\sigma^2_y} - \log \frac{\sigma^2_x}{\sigma^2_y} -1 \right].
	\end{align}
	Note that $\EE X=\mu$ and $\VV(X)= \sigma^2$ are functionally independent, that is, their values can vary separately.  Thus, if we have a unisensory (visual) distribution $\mathcal{N} (\mu_V,\sigma^2_V)$, a unisensory (auditory) distribution $\mathcal{N} (\mu_A,\sigma^2_A)$, and a bisensory distribution $\mathcal{N} (\mu_{VA},\sigma^2_{VA})$, then
	\[ D(f_{VA}||f_V)=  \frac{1}{2} \left[  \frac{(\mu_{VA}-\mu_V)^2}{\sigma^2_V} + \frac{\sigma^2_{VA}}{\sigma^2_V} - \log \frac{\sigma^2_{VA}}{\sigma^2_V} -1 \right]\]
	depends on both the means and variances of the distributions, and so does \\$\mathrm{CRE}_{KLD} =  \min\{D(f_{VA} || f_V), D(f_{VA} || f_A)\}$. Importantly, even under equality of the means,  CRE remains non-zero.
	In particular, if the bisensory distribution has a larger or a smaller variance compared to the unisensory distributions, this is taken into account in the KLD-based measure of MI.
\end{example}
Because of the symmetry of the normal distribution, this example is again not a realistic one for empirical RT data. The following example, however, is often considered to be of a  plausible shape for RTs.
\begin{example}[Log-Normal]
	The density of the log-normal distribution  of a non-negative random variable $X$ is
	\[ f(x;\mu, \sigma) = \frac{1}{x \sigma \sqrt{2\pi}}\exp\left(- \frac{(\log(x)-\mu)^2}{2 \sigma^2}\right), \]
	with $\mu \in (-\infty, +\infty)$ and $\sigma >0$, abbreviated as $X\sim \mathcal{LN}(\mu_x,\sigma^2_x)$. Moreover,
	\[ \EE X = \exp\left(\mu+ \frac{\sigma^2}{2}\right)  \text{ and }  \VV X= [\exp(\sigma^2) -1] \exp(2\mu +\sigma^2).\]
	For random variables $X\sim \mathcal{LN}(\mu_x,\sigma^2_x)$ and $Y\sim \mathcal{LN}(\mu_y,\sigma^2_y)$ with pdfs $f_x, f_y$, respectively, the KLD can be shown \cite{dittrich2013} to be
	\begin{equation}\label{lognorm_kld}
		D(f_x||f_y)= \log \frac{\sigma_y}{\sigma_x} +\frac{1}{2\sigma_y^2}[(\mu_x-\mu_y)^2 +\sigma_x^2- \sigma_y^2.]
	\end{equation}
	Assume $f_{VA}	, f_V, f_A$ are all log-normal distributions, thus $\mathcal{LN}(\mu_{VA},\sigma^2_{VA})$, $\mathcal{LN}(\mu_V,\sigma^2_V)$, and $\mathcal{LN}(\mu_A,\sigma^2_A)$, respectively. Then, 
	
	\[ D(f_{VA}||f_V)= \log \frac{\sigma_V}{\sigma_{VA}} +\frac{1}{2\sigma_V^2}[(\mu_{VA}-\mu_V)^2 +\sigma_{VA}^2- \sigma_V^2] \]
	and
	\[ D(f_{VA}||f_A)= \log \frac{\sigma_A}{\sigma_{VA}} +\frac{1}{2\sigma_A^2}[(\mu_{VA}-\mu_A)^2 +\sigma_{VA}^2- \sigma_A^2].\]
	Both mean and variance of log-normal random variables are functions of both parameters ($\mu$ and $\sigma$), so they cannot vary independently. Still, \[\mathrm{CRE}_{KLD} =  \min\{D(f_{VA} || f_V), D(f_{VA} || f_A)\}\]  depends on both moments.
	
	Two special cases are of interest as well:
	\begin{enumerate}
		\item $\sigma_{VA}=\sigma_V=\sigma_A=\sigma$:
		\[  \mathrm{CRE}_{KLD} =\frac{\min\{(\mu_{VA}-\mu_V)^2),(\mu_{VA}-\mu_A)^2\}}{\sigma^2};\]
		\item  $\sigma_{VA}=\sigma_V=\sigma_A=\sigma$ and $\mu_V=\mu_A=\mu$:
		
		\[   \mathrm{CRE}_{KLD} = \frac{(\mu_{VA}-\mu)^2}{\sigma^2}\]
	\end{enumerate}
	Even in these restricted cases, the MI measure depends on parameter $\sigma^2$ modulating both mean and variance. 
\end{example}

\begin{figure}[ht]
\includegraphics[scale=0.70]{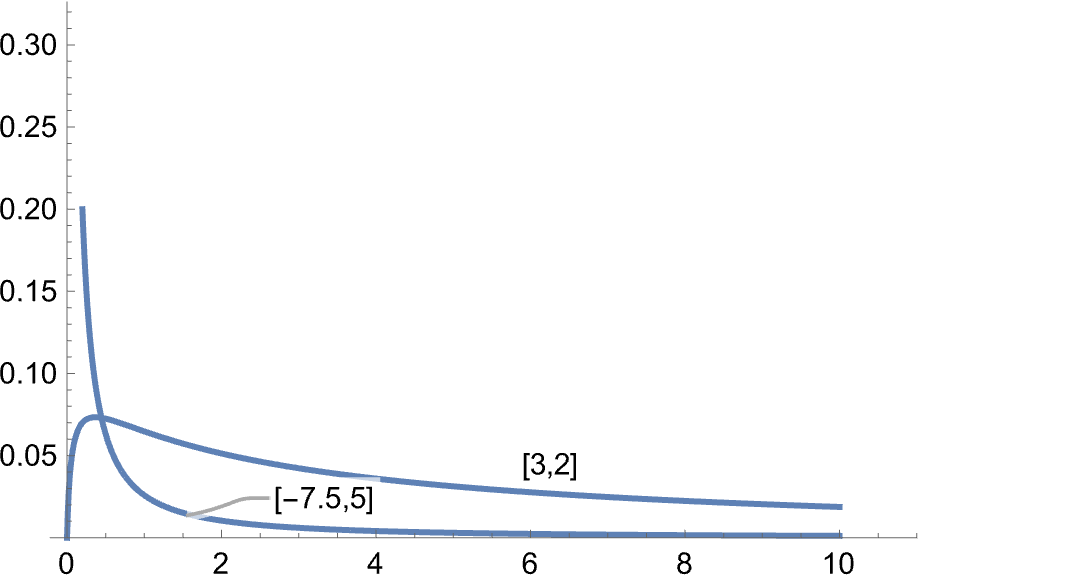}
\caption{Two lognormal probability distribution functions with the same mean~$(\EE=148)$ but different parameters: $(\mu_1 = 3, \sigma_1 = 2)$ vs. $(\mu_2 = -7.5, \sigma_2 = 5)$.}\label{fig1}
\end{figure}

The last example serves to illustrates that measure $\mathrm{CRE}_{KLD}$ is more informative than the traditional one. Indeed, assume that $\EE RT_{VA}
=\EE RT_A$; for convenience, we also assume $\mu_A=\mu_V$ and $\sigma_{A}=\sigma_V$ implying $f_V=f_A$ and $\EE RT_{V}
=\EE RT_A$. Then $\mathrm{CRE}_{RT}=0$ indicates a null effect of multisensory integration. On the other hand, it also means that
\begin{align}
\exp\left[\mu_{VA}+ \frac{\sigma_{VA}^2}{2}\right]&=\exp\left[\mu_{A}+ \frac{\sigma_{A}^2}{2}\right] \;\; \text{or,}\nonumber \\
\mu_{VA}+ \frac{\sigma_{VA}^2}{2}&=\mu_A+ \frac{\sigma_{A}^2}{2},
\end{align}
which does not imply $f_{VA}=f_A$ except if $\mu_{VA}=\mu_{A}$. For example, let $\mu_{VA}=3,\sigma_{VA}=2$ and $\mu_{A}^{(1)}=~ -7.5$ and $\sigma_{A}^{(1)}=~5$. While both pdfs, $f_{VA}$ and $f_A^{(1)}$, have the same mean (148), their shape is very different: $f_A^{(1)}$ has much more probability mass on short values than $f_{VA}$ (see Figure\ref{fig1}). Moreover, by simple calculation 
\begin{equation}
	\mathrm{CRE}_{KLD}= D(f_{VA}||f_A^{(1)})=2.70129,
\end{equation}
For another pdf, $f_A^{(2)}$, with $\mu_{A}^{(2)}=~ - 13$ and $\sigma_{A}^{(2)}=~6$, again with the same mean, we get 
\begin{equation}
	\mathrm{CRE}_{KLD}= D(f_{VA}||f_A^{(2)})=4.20972,
\end{equation}
representing an even larger effect of multisensory integration.

Very similar treatments can be performed with other 2-parameter distributions, e.g., the gamma. Obviously, measure $\mathrm{CRE}_{KLD}$ can also be defined when $f_{VA}, f_V, f_A$ all belong to different distributional families, e.g., log-normal unisensory distributions together with a bisensory Weibull.

\subsection*{KLD-based MI measures based on model predictions}
Besides calculating empirical measures of MI like $\mathrm{CRE}_{RT}$, measures based on models of the integration process are in use as well. Specifically, given a model predicting performance in the crossmodal condition from the unimodal conditions, a KLD-based measure quantifies how ``far away'' the prediction is from the observed data. 
This is analogous to the role of  relative entropy in statistical testing, namely to quantify how much an empirical data set deviates from a hypothesized distribution or model. 

 Let $\tilde{f}_{VA, \,\theta}(t)$ denote the bisensory density predicted by some model with parameter space $\theta \subset \mathbb{R}^d$. The less the observed MI distribution is predictable from the model, the larger the CRE measure should be.  The KLD-based MI measure then is defined by 
\begin{equation}\label{KLDmodel}
	\mathrm{CRE}_{KLD}=  \min_{\theta \subset \mathbb{R}^d}D(f_{VA}||\tilde{f}_{VA,\,\theta}).
\end{equation}
Of course, minimization over the parameter space will be void when a model is parameter-free.

\subsubsection*{The race model: traditional vs. KLD-based MI measures}One of the earliest multisensory models is the (horse) race model, that is, a visual-auditory stimulus complex is supposed to trigger random visual and auditory processes such that the observed RT equals the minimum time of the two, i.e., the 'winner of the race' \cite{raab62}. Thus, combination of the unisensory distributions is here simply defined by probability summation. Under stochastic independence, the bisensory distribution function of the race model is obtained as 
\[ \tilde{F}_{VA}(t) = F_V(t)+F_A(t) -F_V(t)\,F_A(t),\]
with corresponding density
\begin{equation}\label{IND}
\tilde{f}_{VA}(t) = f_V(t) (1-F_A(t))+f_A(t)(1-F_V(t)), \,\; t\ge 0 . \end{equation} 
%
A violation of the race model occurs if the observed distribution $F_{VA}(t)$ is larger than $\tilde{F}_{VA}(t)$ for some $t$. The most traditional MI measure quantifies the amount of violation by defining 
\begin{align}
	R_{VA}^{IND}=\int_{0}^{\infty} [F_{VA}(t)-(F_V(t)+F_A(t)-F_V(t)\,F_A(t))]^+\,\  dt.
\end{align}
Thus, it simply takes the area between the observed bisensory distribution function and the one predicted via the race model\cite{coloniusdiederich2020}. Without assuming stochastic independence, the measure $R_{VA}^{IND}$ can be replaced by the, generally smaller, measure
\begin{align*}
	R_{VA}^{MND}=\int_{0}^{\infty} [F_{VA}(t)-\min\{F_V(t)+F_A(t), 1\}]^+\,\  dt,
\end{align*}
corresponding to maximally negative dependence between the `racers`
\cite{colonius2006}. 
It has been shown that areas  $R_{VA}^{IND}$ and $R_{VA}^{MND}$ are simply equal to the difference between the observed mean (expected value) of the bisensory distribution and the mean predicted by a race model under stochastic independence and maximal negative dependence, respectively\cite{colonius2006}. 
\subsubsection*{KLD-based MI measures for race models} 
%

Under the independent race model  (IND), inserting the  bisensory density into the KLD measure yields 
%
%
\begin{align}\label{indkld}
	\lefteqn{ D(f_{VA}||\tilde{f}_{VA})}\nonumber\\
	& =\int_{0}^{\infty}f_{VA}(t) \log \frac{f_{VA}(t) }{\tilde{f}_{VA}(t) } \,dt   \nonumber  \\
	& =\int_{0}^{\infty}f_{VA}(t) \log \frac{f_{VA}(t) }{f_V(t) (1-F_A(t))+f_A(t)(1-F_V(t))} \,dt 
\end{align}
In this parameter-free form, the integral (\ref{indkld}) can be taken as $ \mathrm{CRE}_{KLD}$. If some specific distributions for the IND model are assumed, minimization of $ D(f_{VA}||\tilde{f}_{VA})$ over the parameter space would be required to compute . 

Comparing the KLD-based measure with traditional one, $R_{VA}^{IND}$, suggests that the former one should be more sensitive with respect to the distributional shapes. The reason is that in the traditional measures, integration is over distributions functions, whereas integration is over densities in KLD measures (\ref{indkld}). Moreover, instead of race models, any other model type (e.g., diffusion co-activation models) predicting $\tilde{f}_{VA}(t)$ can be inserted in $\mathrm{CRE}_{KLD}$. 
\subsubsection*{KLD measures based on mixtures}
This example is special because the model does not predict response enhancement but inhibition of the bisensory RT (for examples, see\cite{welch1986}). Consider a mixture of the unisensory distributions, 
\[\tilde{f}_{VA,\alpha}(t)=\alpha f_V(t) + (1-\alpha) f_A(t),\]
with $0 \le \alpha \le 1$. The cases of $\alpha=1$ or $0$ would yield the components of $\mathrm{CRE}_{KLD}$ again. For a given set of RT distributions, a value of $\alpha$ can be determined that gives the smallest KLD value of $D(f_{VA} || \alpha f_V +(1-\alpha) f_A)$, that is:
\begin{equation}\label{key}
	\alpha^*=\arg \min_{\alpha \in [0,1]} D(f_{VA} || \alpha f_V +(1-\alpha) f_A).
\end{equation}
We then have 
\begin{equation}\label{mixturemodel}
	\mathrm{CRE}_{KLD}=  D(f_{VA}||\tilde{f}_{VA,\,\alpha^*}).
\end{equation}
Note that the value of  $\alpha^*$ may be of interest when interpreted as the relative weight given to the visual component in approximating the bisensory distribution. 
\subsection*{Concluding remarks}
Relative entropy (\textit{aka} Kullback-Leibler divergence, KLD) is an oriented (i.e., non-symmetric) measure of ``distance'' between probability distributions. Here we demonstrate that it can be used to define measures of crossmodal response enhancement that more fully uncover the information about the integration process than classic measures based solely on means or medians of RT data. These novel measures are defined by the relative entropy between some  combination of the unimodal RT distributions and the (observed) crossmodal RT distribution, thus gauging the ``distance'' between the former and the latter. We present several examples where the classic measures are (close to) zero, because the difference between uni-and crossmodal means is (close to) zero whereas the KLD based measure is sensitive to changes in the shape of the RT distributions. While we limit our presentation to theoretical examples, an extension of the approach to empirical data is straightforward, drawing upon the ubiquitous applications of relative entropy in various areas of statistics and machine learning.

Whereas our focus here was on measuring multisensory effects on reaction time, the relative entropy approach could easily be extended to the realm of neuronal data. Note that the most widely used descriptive measure of the magnitude of multisensory integration, measured by absolute spike frequency, is defined as 
\begin{equation}\label{cre}
\mathrm{CRE}=\frac{\mathrm{CM}-\mathrm{SM_{max}}}{\mathrm{SM_{max}}}\times 100,
\end{equation}
where, at the sample level, $\mathrm{CM}$ is the mean absolute number of spikes in response to the crossmodal stimulus and $\mathrm{SM_{max}}$ is the mean absolute number of spikes to the most effective modality-specific component stimulus \cite{meredithstein83}. We have previously suggested to replace (\ref{cre}) by a measure taking into account possible stochastic dependency between the sensory channels under crossmodal stimulation.\cite{Colonius2017}. Using KLD measures on the (theoretical or empirical) spike count frequency distributions, in analogy to the one suggested for RTs here, would go one step further in maximizing the amount of information uncovered by numerical measures of MI.

 \bibliography{colonius_cortex}

\section*{Acknowledgments}
This research was supported by grants from German Science Foundation (DFG): HC by grant CO 94/8-1, AD by grant DI 506/19-1. Computations were performed using Mathematica$^\copyright$(13.3).
\section*{Author contributions statement}
HC and AD wrote the main manuscript text and prepared the figure. Both authors reviewed the manuscript.
\section*{Additional information}
%
\textbf{Competing financial interests} The authors declare no competing financial interest.

\end{document}